# Site selection in single-molecule junction for highly reproducible molecular electronics

Satoshi Kaneko[1], Daigo Murai[1], Santiago Marqués-González[1], Hisao Nakamura[2*], Yuki Komoto[1], Shintaro Fujii[1], Tomoaki Nishino[1], Katsuyoshi Ikeda[3], Kazuhito Tsukagoshi[4*], Manabu Kiguchi[1*]

[1]Department of Chemistry, Graduate School of Science and Engineering, Tokyo Institute of Technology, 2-12-1 W4-10 Ookayama, Meguro-ku, Tokyo 152-8511, Japan, [2]Nanosystem Research Institute (NRI) 'RICS', National Institute of Advanced Industrial Science and Technology (AIST), Central 2, Umezono 1-1-1, Tsukuba, Ibaraki 305-8568, Japan, [3]Graduate School of Engineering, Nagoya Institute of Technology, Gokiso, Showa, Nagoya 466-8555, Japan, [4]WPI Center for Materials Nanoarchitechtonics (WPI-MANA), National Institute for Materials Science, Tsukuba, Ibaraki 305-0044, Japan. *e-mail: kiguti@chem.titech.ac.jp; TSUKAGOSHI.Kazuhito@nims.go.jp; hs-nakamura@aist.go.jp

**Abstract:** Adsorption sites of molecules critically determine the electric/photonic properties and its stability of heterogeneous molecule-metal interfaces. Then, selectivity of adsorption site is essential for development of the fields including organic electronics, catalysis, and biology. However, due to current technical limitations, site-selectivity remains a major challenge because of difficulty in precise selection of meaningful one among the sites. Here, we report the single site-selection at a single-molecule junction by performing newly developed hybrid technique: simultaneous characterization of surface enhanced Raman scattering (SERS) and current-voltage (*I-V*) measurements. The *I-V* response of 1,4-benzenedithiol junctions, reveals the existence of three meta-stable states arising from different adsorption sites. Notably, correlated SERS measurements show selectivity towards one of the adsorption sites "bridge sites". This site-selectivity represents an essential step towards the reliable integration of individual molecules on metallic surfaces. Furthermore, the hybrid spectro-electric technique reveals the dependence of the SERS intensity on the strength of the molecule-metal interaction, showing the interdependence between the optical and electronic properties in single-molecule junctions.

Molecular adsorption sites on the surfaces of metallic and semiconductor materials often govern crucial charge and mass transfer processes at the interface[1-3]. Consequently, a significant portion of the scientific community have recently turned their attention to the role of molecular adsorption sites in: electronics; sensors; energy harvesting and storage; catalysis; biochemistry; and medicine[4-7]. For molecular electronics, where individual molecules are used to mimic conventional electronic components, molecular adsorption sites have a great impact in the overall performance of a molecular component[8-11]. Thanks to the multidisciplinary effort, some impressive milestones have been reached in only four decades since the advent of the field: visualization and manipulation of individual molecules[12-18] and the development of molecular systems with advanced electronic functionality, including diodes, switches, and memories[8,19-23]. However, ungoverned factors like the molecular adsorption site are largely responsible for the characteristic variability in the electrical response of individual molecules[9,10], which ultimately hinders the development of the molecular electronics. Hence, governing the molecular adsorption site *i.e.* site-selectivity, is the key to fully characterize the molecule-metal interface and tackle the significant variability issues. In that sense, single-molecule studies in which an individual molecule is wired between two metallic electrodes represent an elegant and effective approach for the disentanglement of the factors contributing to the averaged response arising from the multiple adsorption sites.

Here, we report the site-selection of a 1,4-benzenedithiol (BDT) single-molecule junction by simultaneous surface-enhanced Raman spectroscopy (SERS)[24-29] and current-voltage (*I-V*)[30,31] measurements, at room temperature. The observed *I-V* response reveals the existence of three meta-stable states arising from different adsorption sites. The SERS signal is selectively detected from only bridge site. Our newly developed hybrid spectro-electric technique provides a deeper insight on the structural and electronic details of a single-molecule junction. This novel approach provides valuable information on the number of molecules bridging the electrode nano-gap, chemical identification, molecular adsorption site, strength of the molecule-metal interaction, and orbital alignment. Furthermore, this technique reveals that the enhancement factor of SERS follows a power law dependence on the strength of the molecule-metal interaction, supported by theoretical calculations.

Mechanically controllable break-junction (MCBJ) substrates feature a freestanding Au nano-junction and were prepared through a series of conventional nanofabrication techniques. An insulating $SiO_2$ layer was located between the phosphor bronze and Au electrodes to limit substrate fluorescence (Fig. 1). Molecular junctions were prepared by depositing a drop of a 1,4-benzenedithiol (BDT) solution (1 mM in EtOH) onto the unbroken Au junction, allowing the analyte to self-assemble on the Au surface via thiol anchoring group of Au-sulfur bond. The Au nano-junction was stretched and eventually broken by gradually bending the substrate using a piezoelectrically controlled push-rod. Although the BDT single-molecule junction has been widely studied[11,32], the adsorption site of molecule was uncontrollable because possible adsorption sites randomly appear in the junction in the conventional junction preparation. All measurements were performed at room temperature.

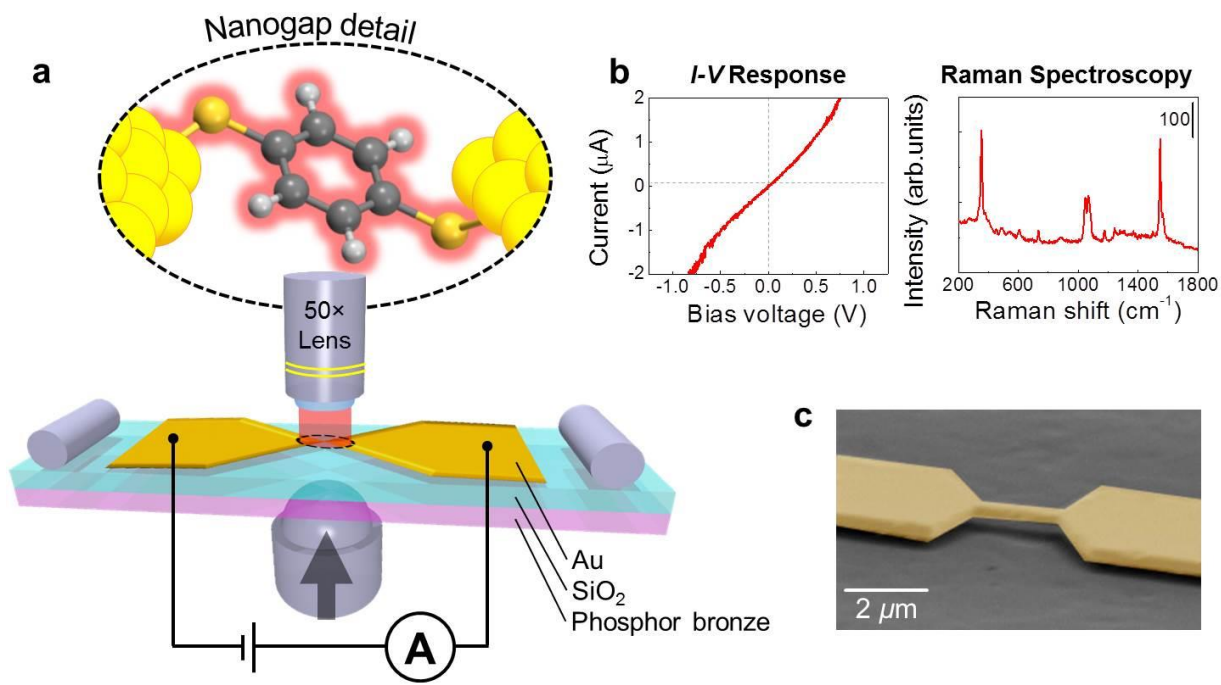

**Figure 1 | Simultaneous SERS and current-voltage measurements of individual molecules. a**, Schematic of original hybrid spectro-electric measurement setup used in this work. The nano-gap detail shows a BDT molecule bridging the gap between the two Au electrodes. An objective lens with 50× magnification was used to focus the near-infrared excitation laser ($λ_{ex}$ = 785 nm, 70 mW) onto the Au nano-junction. Raman spectra were detected using a Raman microprobe spectrometer while simultaneous electrical measurements were performed using a programmable pico-ammeter. Both systems were synchronized and controlled by a custom data flow program. **b,** Typical current-voltage and spectroscopic response obtained for a single molecule suspended between the Au electrodes. **c,** Colored scanning electron microscope image of a freestanding Au nano-junction fabricated via electron-beam lithography, lift-off processes, and isotropic reactive ion etching with $O_2$ plasma. Typical span lengths of the metallic electrode are in the range of ~2 μm.

Upon the bending of the MCBJ substrate in the presence of BDT molecules, three distinct phases (I, II, III) are typically observed in the hybrid spectro-electric data (Fig. 2). Initially, conductance values greater than the fundamental quantum of conductance ($G_0 = 2e^2/h$) are observed for the unbroken Au contact (Fig. 2a, I). Conductance values of approximately 1 $G_0$ indicate the formation of an atomic Au contact[33]. At this stage, the SERS spectra features a number of weak background signals, and the *I-V* measurements exhibit a steep linear response characteristic of a metallic contact (Fig. 2a and c, I). Further bending of the MCBJ substrate leads to a sudden conductance drop, associated with the rupture of the Au junction (Fig. 2a, II). At this point, a conductance step is typically observed at approximately $10^{-2}$ $G_0$. This value is in good agreement with the previously reported conductance values of a BDT single-molecule junction[32], indicating the presence of a single BDT molecule suspended across the electrode nano-gap. Here, a marked enhancement of the SERS intensity and the nonlinear *I-V* response characteristic of molecular charge transport are registered (Figs. 2b and c II). Finally, rupture of the molecular junction leads to a second conductance drop ($< 10^{-4}$ $G_0$) accompanied by a loss of the marked SERS enhancement and a flat *I-V* response (Fig. 2a-c, III). This process was repeated thousands of times until a statistically relevant data set was obtained, and the appearance of the marked SERS enhancement was closely related to the formation of the single-molecule junction with conductance values of approximately 0.02 $G_0$. No significant spectral change was observed upon changing the working voltage bias in the range of 0-200 mV. In addition, the conductance profile of BDT was not significantly affected by light irradiation.

The SERS spectra observed in phases I and III, corresponding to the metallic contact and the broken contact, showed three distinct Raman bands, which can be assigned to the deformation-coupled C-S stretching mode (~325 cm$^{-1}$, $v_{6a}$), a ring breathing mode (~1065 cm$^{-1}$, $v_1$), and a C=C stretching mode (~1558 cm$^{-1}$, $v_{8a}$). The peak positions of these modes indicate that the BDT molecules were adsorbed on Au with one of two anchors (adsorbed-BDT). In phase II, *i.e.*, the single-molecule regime, two additional peaks are observed at approximately 1040 and 1545 cm$^{-1}$, which can be ascribed to the $v_1$ and $v_{8a}$ modes of a BDT molecule bound to Au at both ends (wired-BDT), respectively. The SERS signals in phase II are emitted from a BDT single-molecule junction, whereas the common signals in all phases are from a number of BDTs adsorbed in the vicinity of the nano-junction. The single-molecule origin is also supported by the characteristic variability of SERS with time and within samples. Thermal motion induces fluctuations on the local molecular environment causing variations on the energy and intensity of the vibrational modes. While this effect is generally averaged in the bulk, it gives rise to pronounced variations in peak position and intensity in single-molecule spectroscopy[34]. In the present study, the vibrational energy fluctuation is found be under 20 cm$^{-1}$, which is a two-fold increase with respect to the variability observed in the vibrational spectra of self-assembled monolayers (SAM) of BDT on Au.

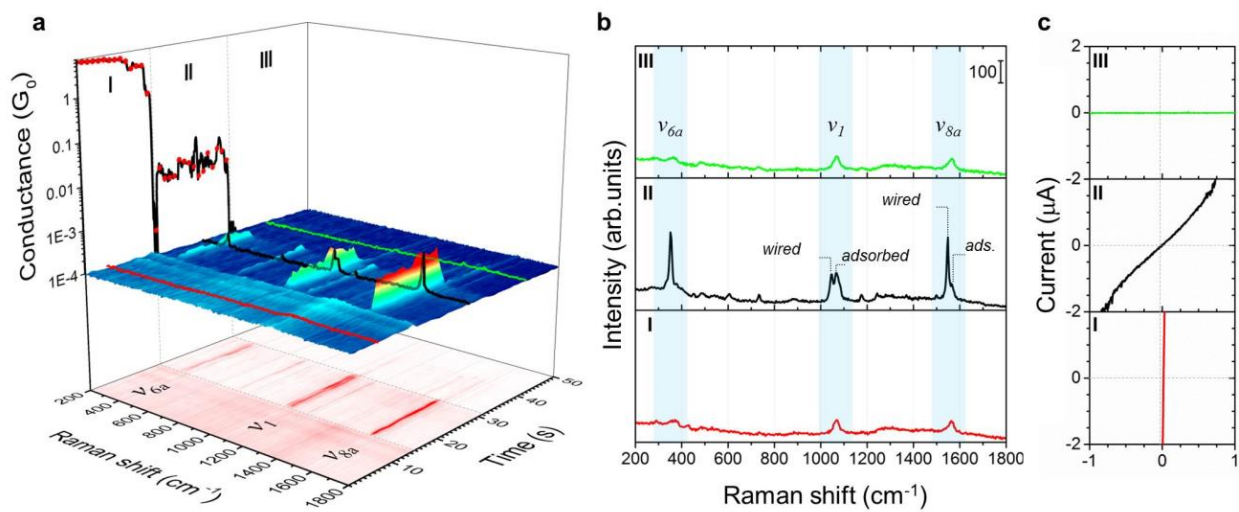

**Figure 2 | Data obtained during the formation and rupture of a single-molecule junction. a**, Three-dimensional representation of the temporal evolution of SERS and conductance measurements upon rupture of the Au contact. Red color corresponds to more intense areas. Three distinct regions can be distinguished: *unbroken metallic contact* (I), featuring conductance values over 1 $G_0$ with weak background SERS signals; the *single-molecule regime* (II), marked by the presence of a conductance step in the region of 10$^{-2}$ $G_0$ and a large enhancement of the

SERS signals corresponding to the $v_{6a}$, $v_1$ and $v_{8a}$ vibrational modes of BDT; and the *broken contact regime* (III), characterized by a conductance drop below $10^{-4}$ $G_0$ and a loss of the marked SERS intensity. The SERS spectrum of the initial state (*unbroken metallic contact*) was subtracted from the following SERS spectra. **b**, Raman spectra extracted from the three-dimensional plot corresponding to the three different regimes. A marked enhancement of the SERS spectrum is typically observed in the single-molecule regime (II), which represents the vibrational modes of BDT molecules bonded to Au through one (adsorbed-BDT) and both molecular termini (wired-BDT). **c**, *I-V* profiles measured simultaneously during the collection of each SERS spectra, showing the characteristic response of each regime: steep linear *I-V* response (I), non-linear response (II) flat *I-V* response (III).

The *I-V* curves in phase II observed in 203 measurements are overplotted in Fig.3a, showing the three statistically high-probable non-linear curves. The strength of electronic coupling across the molecule-metal interface ($\Gamma$) can be obtained by fitting the *I-V* responses to a single-level tunneling transport model[31]. The statistical distribution of strength $\Gamma$ is plotted in Fig. 3b, indicating the three most strength $\Gamma$: high (0.14 eV: denoted as H), medium (0.052 eV: M) and low (0.014 eV: L). The observed conductance values are 0.02, $3\times10^{-3}$, and $3\times10^{-4}$ $G_0$, which are within range of 0.0001 ~ 0.1 $G_0$. The above three coupling strength $\Gamma$'s can be understood by difference in atomic configuration of molecules at the metal-molecule interface.[11] Trends in the electronic coupling $\Gamma$ as well as the electronic conductance of the three states are well-reproduced in the theoretical calculations. Based on theoretical calculations of the BDT-adsorption-site on Au electrodes[35], the experimentally observed H, M, and L states are assigned to the bridge, hollow, and atop adsorption site geometries, respectively. Stronger Au-S interaction in the BDT junction can be achieved by bonding between $p_{x,y}$ of S and $d_{xz,xy}$ of Au at the bridge and hollow sites with larger coordination numbers, which is in contrast to the inert gas-adsorption-system where bonding between $p_z$ of S and $d_{z^2}$ of Au on a atop site with low coordination number dominates the metal-gas interaction.

In Fig.3b, the $v_1$ vibrational mode ($v_1$-active spectra) with intense SERS emissions, which is over 10 counts per second, is colored in orange. The colored signal selectively appears only in H state for the bridge site in the wired-BDT, corresponding to large $\Gamma >0.1$ eV (Fig. 3b). Moreover, the SERS intensity of the Raman bands was found to increase with $\Gamma$ (Fig. 3c). Figure 3d shows the correlation between the intensity of the SERS signal ($I_s$) and $\Gamma$ on a log-log plot. The observed distribution clearly corresponds to a power law relationship, with $I_s \propto \Gamma^{2.1}$ and $\Gamma^{0.7}$ for $v_1$, and $v_{8a}$, respectively. To our knowledge, this is the first *in situ* study to the correlation between the optical and electronic properties in single-molecule junctions.

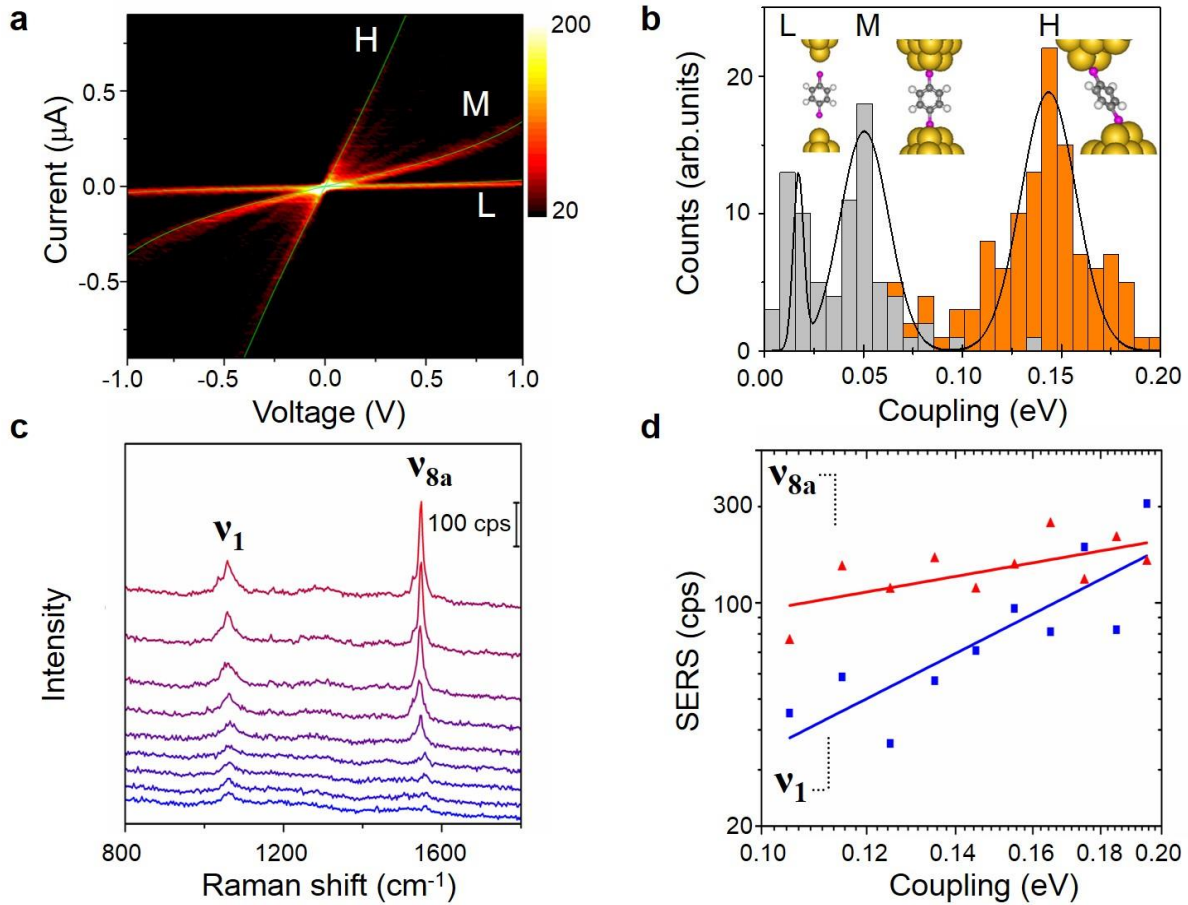

**Figure 3 | Correlated analysis of the optical and electron transport properties of single-molecule BDT junctions. a**, Bi-dimensional *I-V* histogram summarizing the *I-V* response of 203 single-molecule BDT junctions. The three most probable responses were fitted to a single-level tunneling transport model (green). Coupling values ($\Gamma$) of 0.14, 0.052, and 0.014 eV were obtained for the high (H), medium (M), and low (L) profiles, respectively. **b**, Statistical distribution of $\Gamma$ obtained from the individual fitting of 203 single-molecule *I-V* responses and Gaussian fitting (black). Three most probable values are observed at 0.14 (H), 0.052 (M), and 0.014 eV (L), arising from bridge, hollow, and atop molecular adsorption sites, respectively. Orange-colored counts, centered on H, correspond to $v_{8a}$-active samples featuring over 10 CPS. **c**, Single-molecule SERS spectra of BDT showing the intensity enhancement as a function of $\Gamma$: 0.01; 0.02; 0.04; 0.07; 0.12; 0.13; and 0.17 eV. **d**, Correlation between the average intensity of the SERS signal as a function of $\Gamma$ on a log-log plot. The plot includes averaged data from 96 $v_1$- and $v_8$-active samples. The solid line corresponds to the linear least-squares fitting.

The SERS signals gain intensity from two independent contributions: electromagnetic (EM) and chemical (CM) effects[25,26]. The EM effect, which is the major contributing factor, originates from local field enhancement accompanied by the excitation of localized surface plasmon resonances on metallic nanostructures. One of the main sources for the CM effect are charge transfer resonances taking place between metal states near the Fermi level and molecular electronic states. The theoretically expected maximum enhancement factors have been reported to be $10^{10}$ for EM and $10^3$ for CM[25]. Considering the small Raman scattering cross sections of an individual molecule, the enhancement factor on the order of $10^{12}$ is needed to achieve single-molecule detection in SERS[36]. That is, both EM and CM contributions are required in single-molecule SERS. Thus, in the present system, the observed single-molecule SERS signals of wired-BDT should benefit from both EM and CM. In contrast, the SERS signals of adsorbed BDT presumably originate from a number of molecules near the nano-junction where EM enhancement is expected.

When analyzing the charge transfer (CT) mechanism in SERS, two factors have to be taken into consideration, and the energy difference between the molecular orbital (MO) and the Fermi energy of the electrodes ($E_F$) and $\Gamma$[37,38]. In the single-molecule BDT junctions, the energy difference between the lowest unoccupied molecular orbital (LUMO) and the $E_F$ is considerably larger than that of the highest occupied molecular orbital (HOMO) or HOMO-1[39]. The present photon energy, *i.e.*, 1.6 eV (785 nm excitation) may induce CT from the HOMO to an unoccupied metal state (Fig. 4a). Here, the MO with the largest contribution to the CT mechanism is referred to as the "SERS-active MO". It is generally accepted that the HOMO (or HOMO-1) is the dominant conduction orbital in the single-molecule BDT junctions[40,41]. Thus, we consider the dominant conduction MO to be the same as the SERS-active MO.

We evaluated polarizability of a molecule junction based on a single level Anderson model by using a time-dependent nonequilibrium Green's function (NEGF) approach[42]. Figure 4 (b) shows the calculated contour plot of intensity of the SERS signal ($I_s$) of a molecule junction as a function of the MO energy level and $\Gamma$. The blue dotted line at -1.2 eV corresponds to the MO energy in single-molecule BDT junction taken from theoretical prediction[39]. The cross section profile at -1.2 eV plotted in Fig. 4(c) reveals a power law relationship between $I_S$ and $\Gamma$, showing good agreement with our experimental results (Fig. 3(d)). Furthermore, the theoretical profile shows a rapid decay of the $I_S$ for $\Gamma$ below 0.05 eV. This marked decay for small $\Gamma$ accounts for the selective experimental observation of high conductance states in SERS, resulting from the strongly interacting bridge sites among the three possible configurations. Here, we like to note that the molecule-metal coupling is significantly more sensitive to adsorption site and conformational changes than the MO energy level. This streamlined model qualitatively fits the experimentally observed $I_S - \Gamma$ relationship, showing an increased SERS signal for stronger molecule-metal couplings. However, this single-level model cannot fully explain the origin of the different slopes observed in Fig. 3(d) for vibrational modes $v_1$ and $v_{8a}$. A reasonable explanation to the discrepancy in the $I_S - \Gamma$ relationship between $v_1$ and $v_{8a}$ vibrational modes is a multi-orbital contribution to $I_S$. For instance, if both the HOMO and HOMO-1 contribute to CT, but the HOMO-1 is only strongly coupled to one of the electrodes then, electron transport across the junction will still take place exclusively through the HOMO. In that case, a multi-level model has to be adopted for the CT, while the $\Gamma$ values estimated from the $I$-$V$ data will only reflect the HOMO contribution.

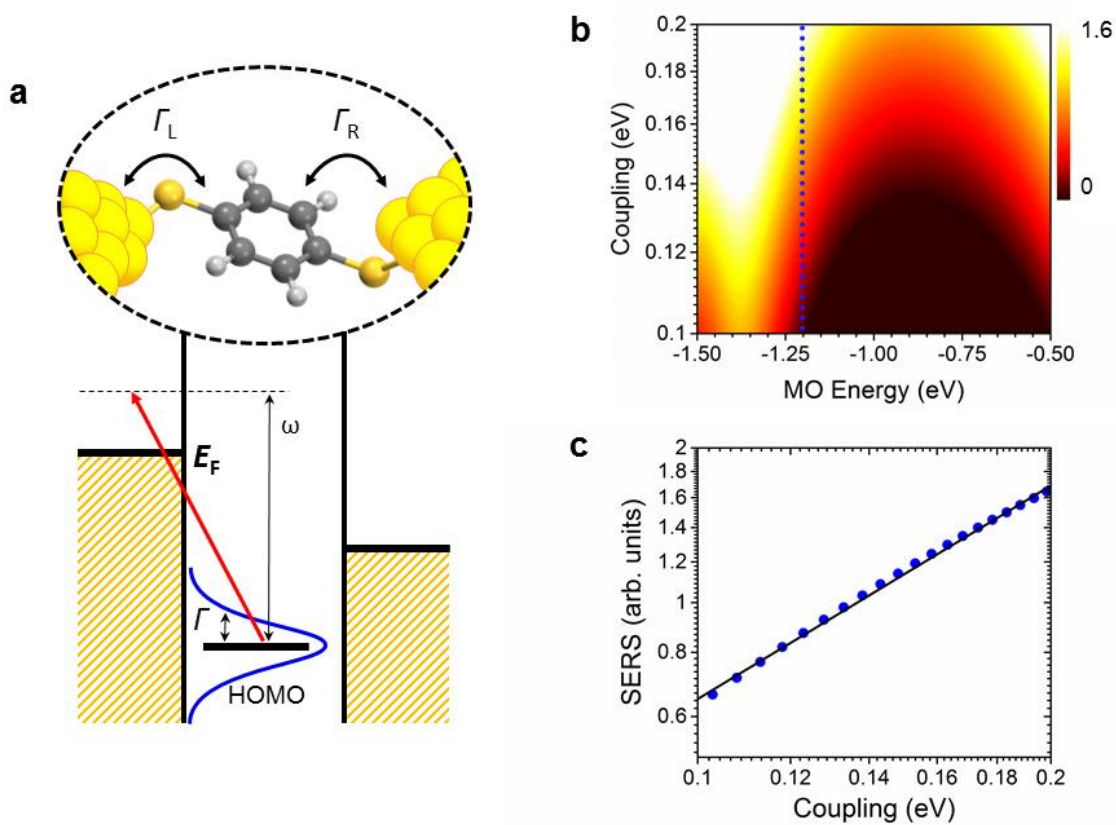

**Figure 4 | Schematic energy diagram and theoretical analysis of correlation profiles for SERS enhancement via the charge transfer resonance mechanism.** Based on the model diagram, the molecule-metal coupling was estimated by theoretical approach. **a,** The transition

probability from the HOMO to the metal unoccupied state is enhanced with an increase in the strength of the molecule-metal interaction. The discrete molecular level is broadened by the interaction, and strong molecule-metal coupling ($\Gamma$) leads to the high conductance of the single-molecule junction. **b,** Contour plot of the intensity of the SERS signal of the single-molecule junction as a function of $\Gamma$ and MO energy level (with $E_F$ = 0 eV). The incident photon, electron-vibron coupling strength and CT dipole constant are set to 1.6 eV, 2.0, 2.0, respectively. The blue dotted line marks the -1.2 eV[38] MO energy cross section, corresponding to single-molecule BDT junction. **c,** The intensity of the SERS signal ($I_s$) as a function of $\Gamma$ on a log-log plot. The linear fitting corresponds to a power law relationship $I_s \propto \Gamma^{1.4}$.

In conclusion, we have introduced a spectro-electric hybrid technique that unveils the bonding site configuration and electronic characterization of individual molecules interacting with metal electrodes at room temperature. The technique allows us to the select solo site among the molecules randomly configured in nano-scale metallic junction. This is the first adsorption site-selective technique, with the SERS signal being selectively enhanced for bridge sites. Site-sensitivity represents a crucial step toward the reliable integration of millions of molecular components into a working device. In addition, this is the first report to provide *in situ* experimental evidence between the molecule-metal interaction and SERS enhancement. It is generally difficult to disentangle the EM and CM contributions to SERS, and to evaluate the strength of the molecule-metal interaction in discussing the CM effect. Our newly developed hybrid technique provides $\Gamma$ which is the strength of the molecule-metal interaction, and the correlation between the molecule-metal interaction and SERS enhancement, that means, the correlation between the electronic and optical properties. The single-molecule junction and the hybrid technique provide a unique platform for the disentanglement of the EM and CM enhancement mechanisms. These findings promote further understanding of the SERS enhancement mechanism and should contribute to the development of SERS technique.

## Methods

**Nanofabrication of the MCBJ substrates.** The substrates used for the mechanically controllable break-junction (MCBJ) technique were prepared through a series of standard nanofabrication techniques. An insulating $SiO_2$ film (~500 nm) was deposited on the polished phosphor bronze substrate of thickness $t = 0.5$ mm by means of sputtering. The $SiO_2$ layer atop the phosphor-bronze substrate provides electrical insulation and limits the intensity of the Raman background scattering. Over the oxide, a polyimide film was deposited by means of spin-coating. The nanosized Au junctions, with the size of the narrowest constriction ~150 nm×120 nm, were prepared atop the polyimide-coated substrate using electron beam lithography and lift-off processing. Metallic layers of Cr and Au (3 nm/130 nm) were thermally evaporated onto the substrate. Subsequently, the polyimide underneath the Au junctions was removed by isotropic reactive ion etching using $O_2$ plasma (80 W) resulting on a free-standing Au nano electrode. Typical span lengths of the junction are in the range of ~2 μm.

**Single-molecule junction preparation.** The nano-fabricated MCBJ substrate was mounted on a three-point bending mechanism, consisting of a stacked piezo-element (NEC tokin) and two fixed counter supports. Molecular junctions were prepared by depositing a drop of a 1,4-benzenedithiol (BDT) solution (1 mM in EtOH) onto the unbroken Au junction allowing molecular self-assembly on the Au surface. The Au nano electrode was stretched and eventually broken by gradually bending the substrate using a piezoelectric push-rod. By retracting the push-rod, the substrate flexing is reduced and the metallic contact can be re-established. For the simultaneous SERS and electrical measurements a self-breaking process was employed. In this case, after the initial rupture of the Au nano-junction, the metallic contact is re-established by retracting the pus-rod until a conductance value of 3 $G_0$ is obtained. The push-rod is hold in position, and the electrical and Raman signals are constantly monitored while the metallic contact is allowed to brake spontaneously from thermal fluctuations and current-induced forces. This immobile substrate methodology is especially appropriate to perform reliable SERS measurements.

***I-V* and SERS measurements.** After self-assembly of the BDT molecules and solvent evaporation, *I-V* and SERS measurements were performed in air at room temperature. The electrical measurements were performed using a Keithley 428 programmable amplifier. Raman spectra were collected using a NanoFinder30 Raman microprobe spectrometer (Tokyo Instruments) employing 1 s integration time. A near-infrared laser ($\lambda_{ex}$ = 785 nm, 70 mW) was used as excitation light. The laser beam was focused onto the Au nano-junction using an objective lens with 50× magnification and 0.95 numerical aperture. The estimated spot size of irradiation was about 1 μm. Once the irradiation area is setup, no further adjustments in the laser position are required during the self-breaking process.

**Acknowledgements**

We thank J.M. van Ruitenbeek for fruitful discussion. This work was financially supported by Grants-in-Aids for Scientific Research in Innovative Areas (26102013, 2511008) and Grant-in-Aid for Scientific Research (A) (No. 21340074) from the Ministry of Education, Culture, Sports, Science and Technology (MEXT), Asahi glass, and the Murata foundation.